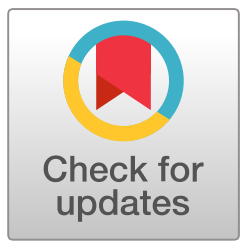

# When a Man Says He Is Pregnant: Event-related Potential Evidence for a Rational Account of Speaker-contextualized Language Comprehension

**Hanlin Wu and Zhenguang G. Cai**

## Abstract

■ Spoken language is often, if not always, understood in a context formed by the identity of the speaker. For example, we can easily make sense of an utterance such as "I'm going to have a *manicure* this weekend" or "The first time I got *pregnant* I had a hard time" when spoken by a woman, but it would be harder to understand when it is spoken by a man. Previous ERP studies have shown mixed results regarding the neurophysiological responses to such speaker–content mismatches, with some reporting an N400 effect and others a P600 effect. In an EEG experiment involving 64 participants, we used social and biological mismatches as test cases to demonstrate how these distinct ERP patterns reflect different aspects of rational inference. We showed that when the mismatch involves social stereotypes (e.g., men getting a manicure), listeners can arrive at a "literal" interpretation by integrating the content with their social knowledge, though this integration requires additional effort due to stereotype violations—resulting in an N400 effect. In contrast, when the mismatch involves biological knowledge (e.g., men getting pregnant), a "literal" interpretation becomes highly implausible or impossible, leading listeners to treat the input as potentially containing errors and engage in correction processes—resulting in a P600 effect. Supporting this rational inference framework, we found that the social N400 effect decreased as a function of the listener's personality trait of openness (as more open-minded individuals maintain more flexible social expectations), while the biological P600 effect remained robust (as biological constraints are recognized regardless of individual personalities). Our findings help to reconcile empirical inconsistencies and reveal how rational inference shapes speaker-contextualized language comprehension. We demonstrate how listeners flexibly adapt its processing strategy based on contextual information, which may be part of the general information-processing principles of the human brain. ■

## INTRODUCTION

A fundamental aspect of spoken language is its dual nature: It carries the linguistic content that conveys the sentence meaning and also the extralinguistic cues that often reveal the speaker's identity (Scott, 2019). For example, listeners can easily make sense of sentences such as "I'm going to have a manicure this weekend" and "The first time I got pregnant I had a hard time" when spoken by a woman; however, it would be harder for them to understand if the same sentences are spoken by a man, as the message of a man getting a manicure violates people's social stereotypical understanding and the message of a man getting pregnant violates people's biological knowledge. These examples highlight how language must be understood in a broader context that involves the identity of the speaker to achieve successful comprehension, but the mechanism is insufficiently explored. In this article, we investigate whether listeners rationally consider the speaker context during real-time language comprehension.

The Chinese University of Hong Kong

### Speaker Identity as Context in Language Comprehension

It has been established that the speaker's identity forms an important context in which language is comprehended (Wu & Cai, 2024; Brown-Schmidt, Yoon, & Ryskin, 2015). This context influences language comprehension by providing information regarding the speaker's sociodemographic background (Creel & Bregman, 2011), including biological gender (e.g., Lattner & Friederici, 2003), age (e.g., Wu, Duan, & Cai, 2024; Kim, 2016; Walker & Hay, 2011), socioeconomic status (e.g., Van Berkum, van den Brink, Tesink, Kos, & Hagoort, 2008), and language background (e.g., Cai, Zhao, & Pickering, 2022; Cai et al., 2017; Martin, Garcia, Potter, Melinger, & Costa, 2016).

Studies show that a speaker's identity, as evidenced by their dialectal accents, modulates listeners' interpretation of word meaning. Cai et al. (2017) showed that listeners had more access to the American meaning of cross-dialectally ambiguous words when spoken by an American English speaker than by a British English speaker. For example, a word such as "bonnet" is more likely to be interpreted as a part of vehicles when spoken in a British accent, but as a type of hat in an American accent (see also Cai, 2022). Similarly, an EEG study by Martin et al. (2016) found that participants showed greater comprehension difficulty—reflected in larger EEG deflections—when words mismatched the speaker's sociodemographic identity indicated by their accent (e.g., British words spoken in an American accent). Similar effects have been found in the recognition of individual words (e.g., Wu et al., 2024; Kim, 2016; Walker & Hay, 2011) as well as the comprehension of sentence meaning (e.g., Van Berkum et al., 2008; Lattner & Friederici, 2003).

These contextual effects of speaker identity have been attributed to the existence of a "speaker model"—a mental model that listeners construct to capture the speaker's characteristics. This model incorporates the listener's beliefs and knowledge about the speaker's age, gender, socioeconomic status, and other characteristics used to interpret utterances. Cai et al. (2017) found that the speaker context effect did not depend on the accentedness of a word token. Listeners still had more access to the American meaning of word tokens that were morphed to be accent-neutral as long as they believed that the word tokens were produced by an American English speaker. This suggests that the contextual effect of speaker identity originates from a higher-level model of the speaker rather than surface acoustic details. The speaker model account gains further support from studies showing that language comprehension can be influenced by a prior introduction to the speaker's identity (e.g., Johnson, Strand, & D'Imperio, 1999) or simply a photo of the speaker (e.g., Hernández-Gutierrez et al., 2021; Hay, Warren, & Drager, 2006).

### N400 and P600 in Speaker-contextualized Language Comprehension

Over the past decades, research using EEG has explored how language is comprehended in a speaker context—a process we term *speaker-contextualized language comprehension*. A significant body of research focused on the N400 and P600 ERP components (there are also other components such as late frontal positivity; see Ryskin, Ng, Mimnaugh, Brown-Schmidt, & Federmeier, 2020). In an early study, Lattner and Friederici (2003) presented participants with implausible self-referent utterances where the linguistic content violated people's gender-based expectations—for example, "I like to play *soccer*" spoken by a woman versus a man. They found that the critical word "soccer" elicited a larger P600 deflection when spoken by a woman, demonstrating the influence of the social stereotypes that associate soccer more strongly with men than women. Van Berkum et al. (2008) expanded on this by including contrasts of age and socioeconomic status alongside gender. Using sentences such as "Every evening I drink some *wine* before I go to sleep" spoken by either a child or an adult, they found that the critical word "wine" elicited a larger N400 deflection, rather than P600, when spoken by a child. They attributed this N400 effect to early integration of the linguistic content with the speaker's identity and suggested that the P600 effect in Lattner and Friederici's study resulted from suboptimal stimulus control.

The N400 effect reported by Van Berkum et al. (2008) has been replicated in studies examining mismatches between linguistic content and speaker context, particularly involving social stereotype violations (Pélissier & Ferragne, 2022; Martin et al., 2016; van den Brink et al., 2012). It has also been supported by studies showing that the N400, reflecting lexical-semantic processing, is modulated by the speaker context (Wu et al., 2024; Foucart & Hartsuiker, 2021; Grant, Grey, & van Hell, 2020; Brothers, Dave, Hoversten, Traxler, & Swaab, 2019; Foucart, Santamaría-García, & Hartsuiker, 2019; Bornkessel-Schlesewsky, Krauspenhaar, & Schlesewsky, 2013) or by whether the speaker context is available or not (Hernández-Gutierrez et al., 2021).

However, some studies report a P600 effect instead of N400 in response to speaker–content mismatch. Using Van Berkum's paradigm, Foucart et al. (2015) found P600 (also labeled as "late positive potential") when the linguistic content mismatched the speaker context. Similarly, van den Brink et al. (2012) observed N400 early in the experiment but P600 later on. Even Van Berkum et al. (2008) noted that, while their study showed an overall N400 effect, gender contrasts specifically elicited an additional P600 effect. Further studies show that the speaker context modulates P600 effects related to grammatical (Caffarra, Wolpert, Scarinci, & Mancini, 2020; Zhou, Garnsey, & Christianson, 2019; Hanulíková & Carreiras, 2015; Hanulíková, van Alphen, van Goch, & Weber, 2012) and lexical-semantic processing (Foucart et al., 2019; Regel, Coulson, & Gunter, 2010).

Overall, the literature presents inconsistent findings regarding the N400 and P600 effects when the speaker's identity mismatches the content of their utterances. The implications of these patterns for how speaker information is integrated remain the focus of ongoing empirical research and theoretical discussion. While differing in the time course, both N400 and P600 effects have been interpreted as indexing the cross-domain integration of the speaker context and linguistic content. Some research interpreted the observation of N400 effects as reflecting early-stage integration, with speaker contexts being processed simultaneously with sentence semantics (Pélissier & Ferragne, 2022; Martin et al., 2016; van den Brink et al., 2012; Van Berkum et al., 2008). Conversely, P600 effects were suggested to reflect later-stage integration, where speaker identities are processed after the construction of speaker-independent sentence meaning (Caffarra et al., 2020; Foucart et al., 2015; Lattner & Friederici, 2003). In this article, however, we propose that the N400 and P600 effects in speaker-contextualized language comprehension may not reflect the distinction between earlier and later integration but rather indicate a mechanism of rational inference through which comprehenders work to achieve speaker-congruent interpretation.

This rational inference perspective views language comprehension as an active process of probabilistic inference. It proposes that listeners integrate incoming linguistic signals with prior knowledge—including linguistic structures, world knowledge, and discourse context—to arrive at the most probable interpretation of a sentence. Crucially, this inferential process also involves an implicit consideration of the "noisy channel" nature of communication (Gibson, Bergen, & Piantadosi, 2013; Levy, 2008), meaning that listeners may assess the likelihood that an unexpected or implausible input could have arisen from an error (noise), either on the part of the speaker (e.g., a slip of the tongue) or in their own perception (e.g., mishearing). As we will discuss next, within such a framework, ERP components like N400 and P600 are increasingly understood to reflect different facets of this ongoing inferential process, with N400 reflecting the integration of the current input with prior context and P600 having a role in monitoring and correcting potential errors.

## N400 and P600 in Rational Language Processing

The N400 and P600 are two ERP components consistently linked to language processing. The N400, a negative-going deflection peaking around 400 msec after stimulus onset, is traditionally associated with semantic violations (Kutas & Hillyard, 1980, 1984). For example, in "I take my coffee with cream and *dog*," the word "dog" elicits a larger N400 compared with the expected word "sugar" because it violates semantic constraints established by the preceding context. Subsequent research has demonstrated that the N400 is not simply an index of semantic violations but rather reflects a broader process of integrating word meanings with both semantic context and world knowledge. Hagoort, Hald, Bastiaansen, and Petersson (2004) demonstrated this by comparing semantic violations (e.g., "Dutch trains are *sour* and very crowded") with world knowledge violations (e.g., "Dutch trains are *white* and very crowded," when Dutch trains are characteristically yellow). Both violations elicited similar N400 effects in onset, peak latency, amplitude, and topographic distribution, suggesting parallel integration of lexical-semantic and general world knowledge during language comprehension (Hagoort, Baggio, & Willems, 2009; Hagoort et al., 2004). This integration view gains further support from studies showing similar N400 modulation by violations of discourse contexts (Nieuwland & Van Berkum, 2006), and co-speech gestures (Willems & Hagoort, 2007) or pictures (Willems, Özyürek, & Hagoort, 2008). These findings suggest that various sources of contextual information—linguistic and nonlinguistic—are all integrated simultaneously during language comprehension.

The P600, a positive-going deflection peaking around 600–1000 msec after stimulus onset, was initially linked to the processing of syntactic errors (Osterhout & Holcomb, 1992). For example, early studies found that syntactic errors like subject–verb agreement errors (e.g., "The spoiled child *throw* the toys on the floor") consistently elicited P600 effects (Hagoort, Brown, & Groothusen, 1993). Subsequent research revealed the P600's broader role in processing various errors, including spelling mistakes (Münte, Heinze, Matzke, Wieringa, & Johannes, 1998), thematic role violations (e.g., "The hearty meal was *devouring* …"; Kim & Osterhout, 2005), and violations of grammatical-semantic constraints on linear order (e.g., "Jennifer rode a gray *huge* elephant"; Kemmerer, Weber-Fox, Price, Zdanczyk, & Way, 2007). These findings suggest that P600 reflects a general process of error detection and reanalysis when input violates expectations and requires interpretation revision (van de Meerendonk, Kolk, Chwilla, & Vissers, 2009).

While the N400 and P600 have traditionally been studied as independent indices, recent research has begun to examine how their dynamic relationship—the relative magnitudes between N400 and P600—reflects rational inference during language comprehension. This new perspective aligns with theories of rational language processing, which propose that comprehension involves probabilistic inference where comprehenders integrate prior expectations with incoming information while considering potential communication channel noise, such as mishearing or speech errors (Levy, 2008).

Consider how people comprehend implausible sentences like "The mother gave the candle the daughter." According to Gibson et al. (2013), comprehenders evaluate such input by considering the likelihood that it has been noise-corrupted from an intended plausible sentence "The mother gave the candle *to* the daughter" through the omission of "to." Similarly, "The mother gave the daughter *to* the candle" might be interpreted as a noise-corrupted version of "The mother gave the daughter the candle" due to the accidental insertion of "to" (Cai et al., 2022).

Recent studies suggest that N400 and P600 reflect different aspects of this rational inference process. Using implausible sentences like those above, studies found that the N400–P600 pattern varies systematically with the likelihood that there is an error in the perceived sentence input (Li & Ettinger, 2023; Ryskin et al., 2021). When encountering an implausible sentence like "The mother gave the daughter to the candle," comprehenders show reduced N400 but enhanced P600, as an attempt to correct the potential error and revise it as a plausible alternative ("The mother gave the daughter the candle"). Conversely, when no explicit error is detected, N400 becomes more prominent, indexing the difficulty of lexical-semantic processing. This pattern suggests that while N400 reflects the rapid integration of prior probabilistic expectations about the intended meanings, P600 indexes the engagement of a correction mechanism when the input is deemed likely to contain an error.

## The Current Study

The rational inference framework—where the N400 reflects integration difficulty and the P600 reflects error correction—may resolve the N400/P600 inconsistencies in the literature for mismatches between linguistic content and speaker identity. Under the rational inference framework, comprehenders not only evaluate the plausibility of the sentence meaning but also evaluate the plausibility of utterances by considering the speaker context: They rationally integrate speaker identity with the linguistic content to arrive at interpretations, considering both their prior knowledge about the speaker and the actual linguistic input they receive.

When encountering speaker–content mismatches, comprehenders may first attempt to arrive at a "literal" understanding of the utterance. For example, hearing "I'm going to have a *manicure* this weekend" from a male speaker (as compared with a female speaker) may challenge social stereotypes but still allow for a straightforward interpretation—the speaker literally plans to get a manicure. While such an interpretation is possible, it may require additional cognitive effort (as compared with when it matches social stereotypes) to integrate this information with prior expectations, resulting in an N400 effect that reflects this integration difficulty.

However, when a "literal" understanding becomes extremely implausible, such as hearing "The first time I got *pregnant* I had a hard time" from a male speaker, comprehenders may need to revise their perception—either of the linguistic content or the speaker identity—to arrive at a more plausible "nonliteral" interpretation (Cai et al., 2022; Gibson et al., 2013). This revision process typically manifests as a P600 effect, reflecting the engagement of an error correction mechanism when the input substantially deviates from prior knowledge.

This framework may explain the apparent inconsistency in previous findings. Studies reporting P600 effects (e.g., Foucart et al., 2015) often included utterances that violated the speaker's biological constraints, such as "Today I am feeling sick, I will need to visit my *pediatrician* again" spoken by an adult compared with a child, or "I have *erection* problems due to stress" spoken by a woman compared with a man. In these cases, the substantial deviation from biological knowledge triggers a process of error correction rather than integration.

Furthermore, if N400 and P600 indeed reflect different aspects of rational inference and, more generally, distinct modes of information processing in the brain, they should be differentially affected by individual differences. We focus on "openness," also known as "openness to experience" (Digman, 1990) or "open-mindedness" (Zhang et al., 2022; Soto & John, 2017)—a trait that negatively correlates with stereotypical thinking (Crawford & Brandt, 2019; Chen & Palmer, 2018; Sibley & Duckitt, 2008; Flynn, 2005). For social mismatches where literal interpretation is possible, we predict that the integration difficulty (as reflected in N400) will be modulated by openness, with less open-minded individuals showing larger N400 effects due to stronger stereotypical expectations. In contrast, for biological mismatches where error correction is necessary, the P600 effect should remain robust across individuals regardless of their openness levels, as biological knowledge is typically learned and understood with a high degree of consistency across individuals.

# METHODS

## Participants

We recruited 64 neurotypical native Mandarin Chinese speakers (32 women and 32 men; mean age = 22.97 years, *SD* = 1.98 years). Four participants were excluded from data analysis due to excessive artifactual contamination (see EEG Recording and Preprocessing section), resulting in a final sample size of 60 participants—a size exceeding those of previous studies (e.g., Pélissier & Ferragne, 2022; Martin et al., 2016; Van Berkum et al., 2008). All participants provided informed consent before the experiment began. The study protocol adhered to Helsinki Declaration ethical standards and was approved by the Joint Chinese University of Hong Kong-New Territories East Cluster Clinical Research Ethics Committee.

## Design

We adopted a 2 (Plausibility: plausible vs. implausible) × 2 (Type: social vs. biological) factorial design. Plausibility was manipulated within participants and within items: A sentence item was plausible or implausible based on whether the linguistic content matches the speaker's gender/age (as indicated by their voice). For most analyses, data from the gender and age dimensions were combined. Type referred to the type of (im)plausibility being social (e.g., a person discussing getting a manicure) or biological (e.g., a person discussing pregnancy) and was manipulated within participants and between items (see Table 1 for examples). To ensure each participant heard only one version (plausible or implausible) of any given sentence item, we created four experimental lists. Each list presented a given item in either its plausible or implausible audio version, with items counterbalanced across lists. Participants were randomly assigned to one of these four lists. The order of trials within each list was randomized for each participant.

## Materials

We constructed 80 Mandarin self-referential sentences (see Table 1 for examples; see full list in Table S1 in the Supplemental Material). These sentences were divided into eight categories of 10 sentences each based on the combination of the Type of linguistic content (social or biological) and the speaker's identity (male, female, adult, or child) with which the linguistic content was designed to be plausible. Specifically, there were four categories of social content sentences (each designed to be plausible with male, female, adult, or child speakers, respectively) and four categories of biological content sentences (similarly designed to be plausible with male, female, adult, or child speakers, respectively).

**Table 1.** Examples of Stimuli with English Translations

| *Category* | *Example (English Translation)* |
|---|---|
| Socially plausible with | |
| male speakers | 在工作单位我一般都是穿*西服*打领带。 |
| | (At the workplace I usually wear a *suit* and a tie.) |
| female speakers | 这个周末我要先去做*美甲*然后理发。 |
| | (This weekend I'm going to get a *manicure* and then a haircut.) |
| adult speakers | 我最近*上班*压力太大需要休息。 |
| | (I've been *working* too hard lately and I need a break.) |
| child speakers | 他把我的*玩具*抢走了我要去找妈妈告状。 |
| | (He took my *toys* away from me and I'm going to tell mummy about it.) |
| Biologically plausible with | |
| male speakers | 我需要定期去医院检查*前列腺*的健康状况。 |
| | (I need to go to the hospital to check my *prostate* on a regular basis.) |
| female speakers | 我第一次*怀孕*的时候过得很艰难。 |
| | (The first time I got *pregnant* I had a hard time.) |
| adult speakers | 我发现我脸上的*老年斑*越来越多了我正在寻找新的治疗方法。 |
| | (I noticed that I'm getting more and more *age spots* on my face and I'm looking for new treatments.) |
| child speakers | 我在等我的*乳牙*掉下来然后我要把它扔到房顶上。 |
| | (I'm waiting for my *milk tooth* to fall out and then I'm going to throw it on the roof.) |

We designed these sentences following a set of rules. First, the speaker-contextualized plausibility always emerged at a critical word (either disyllabic or trisyllabic, italicized in the examples). Second, the critical word was always preceded by at least three syllables (equivalent to three Chinese characters) to ensure that listeners had established the speaker context before encountering the critical word (McAleer, Todorov, & Belin, 2014; Scharinger, Monahan, & Idsardi, 2011). Third, the critical word was always followed by at least three characters before the sentence ended to avoid sentence "wrap-up" effects. In addition to these rules, social and biological sentences were matched on the critical word frequencies, $t(75.50) = 0.03$, $p = .978$, using SUBTLEX-CH (Cai & Brysbaert, 2010) and critical word lengths, $t(77.32) = -0.55$, $p = .582$.

For each target sentence, we created two audio versions using different speaker voices (16 different speakers in total), one rendering the sentence plausible and the other implausible given the speaker's identity. To minimize acoustic differences beyond the intended gender and age manipulations and to ensure high acoustic control, we generated audio files using text-to-speech (TTS), selected for its capacity to produce natural-sounding speech. This approach helped to control for potential confounds often present with human speakers, such as volume, accent, and speech rate. Participants were not informed that the voices were computer-generated, and no participant spontaneously reported perceiving the voices as artificial during or after the experiment. The use of high-quality TTS for manipulating speaker characteristics like age and gender while maintaining perceived naturalness has been validated in similar paradigms (e.g., Wu et al., 2024, who similarly used TTS and demonstrated good naturalness ratings, around 5.33 on a 7-point scale, for stimuli generated for similar experimental purposes). Critical word duration was matched between the two audio versions of each sentence, $t(79) = -0.51$, $p = .610$. Additionally, we included 80 gender- and age-neutral sentences as filler items.

### Procedure

Before the EEG experiment, we conducted a plausibility rating test including 30 participants (15 women and 15 men; mean age = 23.57 years, *SD* = 3.23 years) who did not participate in the subsequent EEG experiment. Participants were individually tested in a laboratory environment in which they listened to each audio sentence and rated how plausible they thought it was for the speaker in the audio to say the sentence on a 7-point Likert scale (1 = *extremely implausible*, 4 = *neutral*, 7 = *absolutely plausible*).

In the EEG experiment, participants were individually tested in a soundproof booth designed for EEG signal acquisition. During the experiment, participants' EEG signals were recorded while they listened to the audio. Each trial began with a fixation cross on the center of the screen for 1000 msec. The audio was then played while the fixation cross remained on the screen until 1000 msec after the audio offset. Each trial was followed by an interval of 3600 msec. To ensure their attentive listening, participants were required to answer a yes/no probe question about the content of the utterance in 50% of the filler trials. For both the rating test and EEG experiment, participants completed the Big Five Inventory-2 (Mandarin version; Zhang et al., 2022), of which the subscore of Openness was later used in the analyses.

### EEG Recording and Preprocessing

The EEG was collected using 128 active sintered Ag/AgCl electrodes positioned according to an extended 10–20 system. All electrodes were referred online to the left earlobe. Signals were recorded using a g.HIamp amplifier and digitalized at a sampling rate of 1200 Hz. All electrode impedances were maintained below 30 kΩ throughout the experiment. EEG data preprocessing was performed using customized scripts and the FieldTrip toolbox (Oostenveld, Fries, Maris, & Schoffelen, 2011) in MATLAB (The MathWorks). The raw EEG data were bandpass-filtered offline at 0.1–30 Hz (Luck, 2014, 2022; Tanner, Morgan-Short, & Luck, 2015), resampled at 500 Hz, and re-referenced to the average of the left and right earlobes (A1 and A2). Independent component analysis was applied to the 1–30 Hz filtered continuous data with dimensionality reduction to 30 components to identify and remove ocular artifacts. The data were then epoched from −200 to 1200 msec relative to the critical word onset and baseline-corrected by subtracting the mean amplitude from −200 to 0 msec relative to the critical word onset. Epochs with amplitudes exceeding ±100 μV were considered to contain artifacts and thus excluded (8.37%). The data of four participants (four men) with more than 40% of trials containing artifacts were excluded, leaving a total of 60 participants (32 women and 28 men; mean age = 22.97 years, *SD* = 2.05 years) for further analyses.

## RESULTS

### Plausibility Rating

Linear mixed-effects (LME) modeling was used on the plausibility rating data of all 30 participants in the rating test. Plausibility (plausible = −0.5, implausible = 0.5) and type (social = −0.5, biological = 0.5) were included as fixed effects. Participant and item were included as random effects (see Table S2 in the Supplemental Material for model structures). For all LME analyses, we used the maximal random-effect structure justified by the data and determined by forward model comparison ($\alpha$ = .2; see Matuschek, Kliegl, Vasishth, Baayen, & Bates, 2017).

As shown in Figure 1, the results showed a significant main effect of Plausibility ($\beta = -3.59$, $SE = 0.19$, $t = -19.15$, $p < .001$), a significant main effect of Type ($\beta = -0.83$, $SE = 0.12$, $t = -6.58$, $p < .001$), and a significant interaction between Plausibility and Type ($\beta = -1.31$, $SE = 0.24$, $t = -5.37$, $p < .001$). Plausible utterances (mean rating = 6.37) generally had a higher rating than implausible ones (mean rating = 2.78). Social utterances (mean rating = 4.99) generally had a higher rating than biological ones (mean rating = 4.16). The absolute size of the implausibility effect, calculated as the rating difference between plausible and implausible utterances, was smaller for social utterances (−2.94) than for biological utterances (−4.25).

To test whether a listener's personality trait of openness could predict their perception of the plausibility of each type of utterance, we additionally included Openness (a scaled continuous variable) as a fixed effect interacting with Plausibility and Type (see Table S2 in the Supplemental Material for model structures). The results showed a significant three-way interaction among Plausibility, Type, and Openness ($\beta = -0.28$, $SE = 0.12$, $t = -2.35$, $p = .026$). Separate analyses of social and biological utterances showed a significant interaction between Plausibility and Openness for social utterances ($\beta = 0.31$, $SE = 0.15$, $t = 2.13$, $p = .043$), suggesting that the difference in plausibility rating between the socially plausible and implausible utterances decreased as a function of a rater's openness score (i.e., raters with higher score seemed to be more tolerant with socially implausible utterances). In contrast, this interaction did not reach statistical significance for biological utterances ($\beta = 0.03$, $SE = 0.18$, $t = 0.18$, $p = .860$), showing no evidence of an impact of openness on the biological implausibility effect (Figure 2).

Overall, ratings indicated that biological implausibility effects were more pronounced than social implausibility effects, and only social implausibility effects were (negatively) predicted by listeners' personality traits of openness.

### EEG Amplitude

We focused our analyses of EEG amplitude on the time windows of 300–600 and 600–1000 msec after the critical word onset, corresponding to the typical time windows of N400 (e.g., Kutas & Hillyard, 1980) and P600 (e.g., Lattner & Friederici, 2003), respectively. LME models were applied to the mean amplitudes within these windows for each trial, as LME methods are considered more robust than traditional ANOVA-based approaches in amplitude analyses (Heise, Mon, & Bowman, 2022; Frömer, Maier, & Abdel Rahman, 2018). We first analyzed the topography of speaker-contextualized implausibility effects, which guided the selection of an ROI for subsequent analyses.

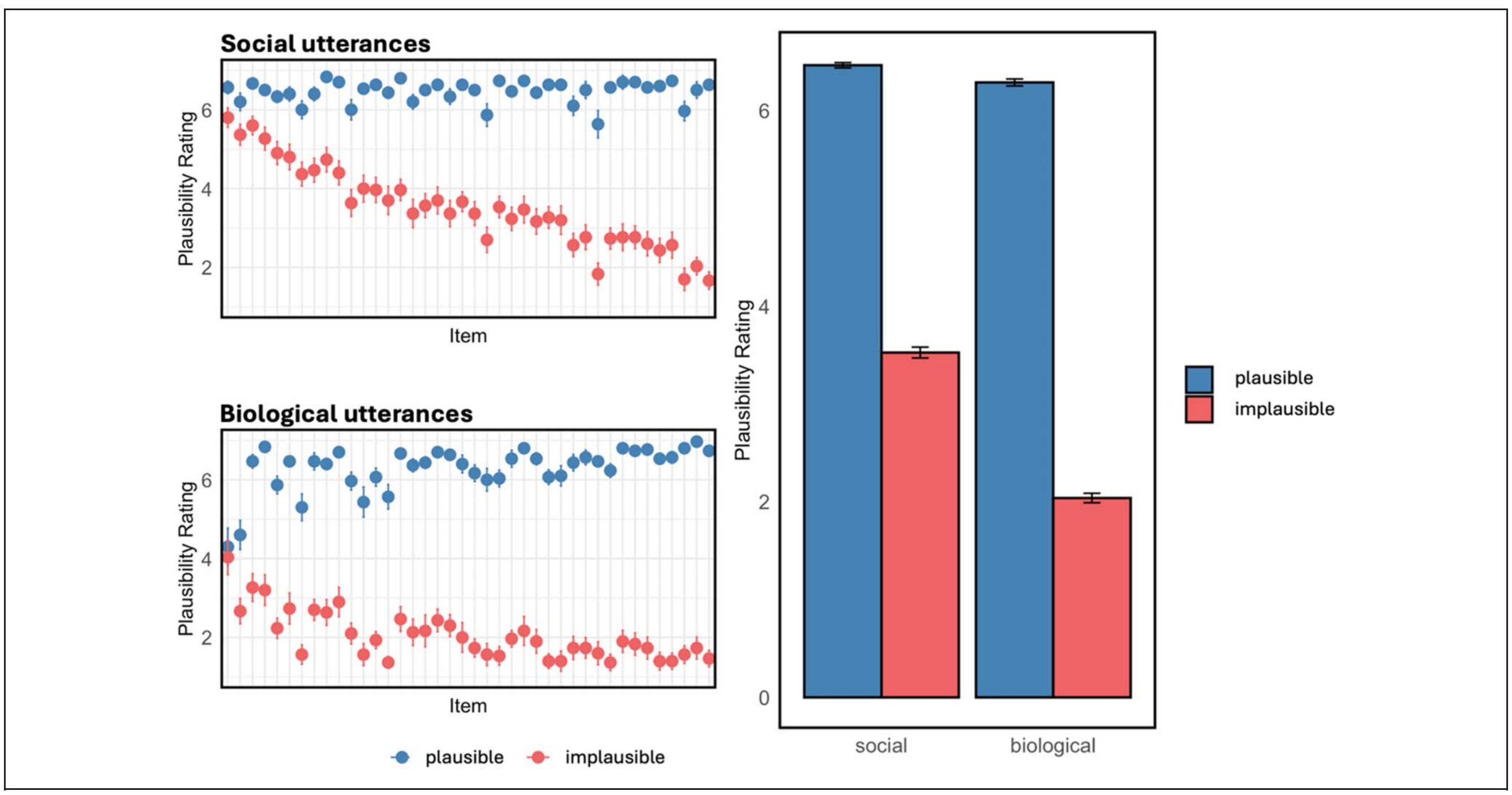

**Figure 1.** Plausibility rating results. The items are ordered according to the absolute size of the implausibility effect.

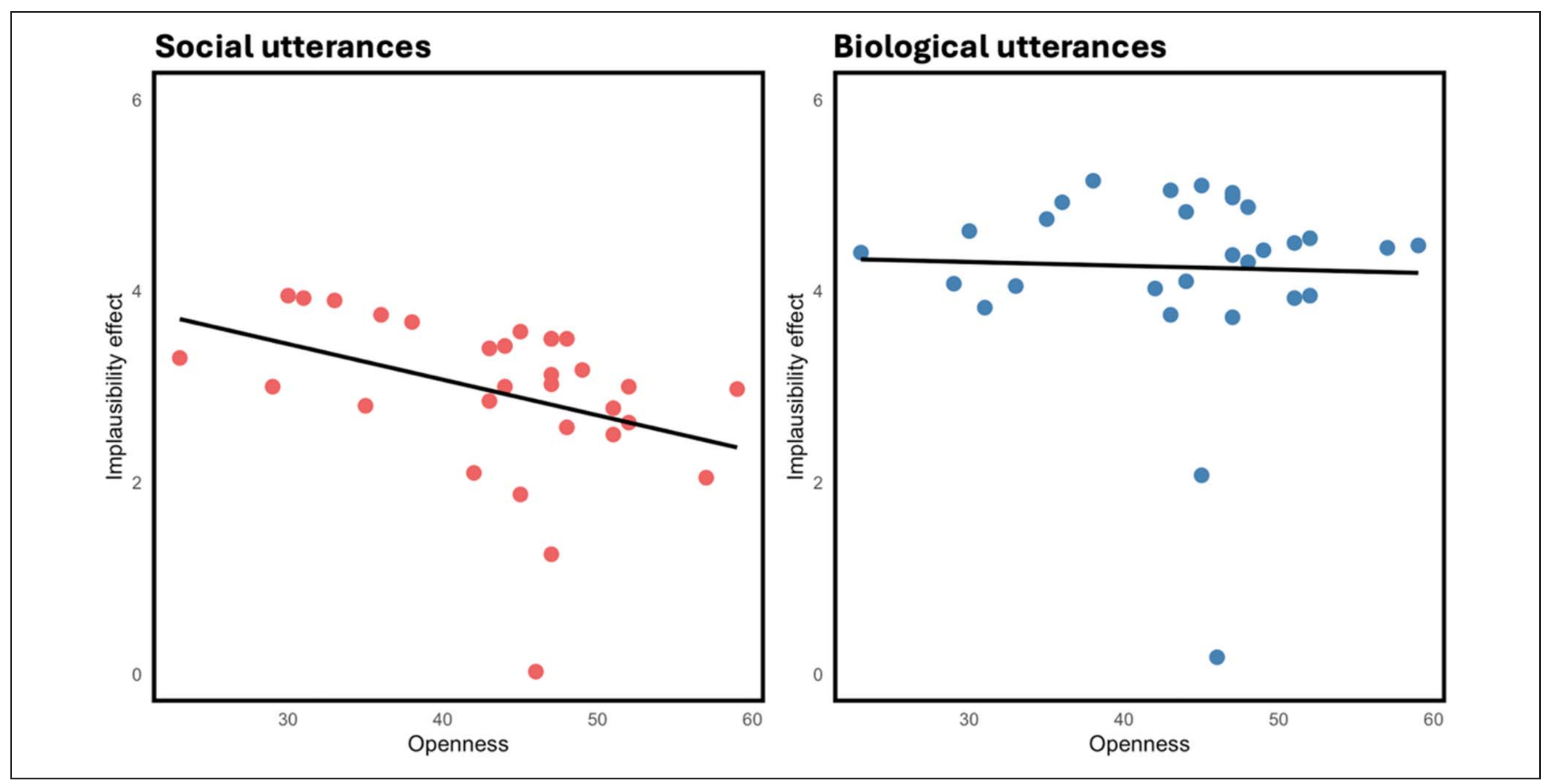

**Figure 2.** Social and biological implausibility effects of plausibility rating predicted by the rater's openness score.

*Topography Analyses*

To explore the topography of speaker-contextualized implausibility effects on EEG amplitudes, we conducted analyses of Posteriority and Laterality (see also Martin et al., 2016, for a similar approach). As depicted in Figure 3A, scalp sites were selected and divided into four regions: left anterior, right anterior, left posterior, and right posterior. Each region consisted of 17 sites (for the full lists, see Table S5 in the Supplemental Material), with the mean amplitudes collapsed across all sites.

To determine whether speaker-contextualized implausibility effects differed in magnitude between anterior and posterior sites, we conducted a posteriority analysis. We fit LME models with Plausibility, Type, and Posteriority (anterior = −0.5, posterior = 0.5) as interacting fixed effects (see Table S3 in the Supplemental Material for model structures). Our focus was on the three-way interaction among Plausibility, Type, and Posteriority. In the time window of 300–600 msec after the critical word onset, there was a marginally significant interaction among Plausibility, Type, and Posteriority ($\beta = 0.91$, $SE = 0.47$, $t = 1.92$, $p = .054$), suggesting a trend of difference in the scalp topographies between social and biological implausibility effects. Separate analyses of social and biological utterances showed a significant interaction between plausibility and posteriority for social utterances ($\beta = -0.66$, $SE = 0.33$, $t = -2.02$, $p = .044$), suggesting larger implausibility effects over posterior sites (−0.90 μV) compared with anterior sites (−0.22 μV). However, this interaction was not significant for biological utterances ($\beta = 0.25$, $SE = 0.34$, $t = 0.74$, $p = .460$), suggesting comparable implausibility effects between posterior (0.22 μV) and anterior sites (−0.02 μV). In the time window of 600–1000 msec after the critical word onset, there was a significant interaction among Plausibility, Type, and Posteriority ($\beta = 1.67$, $SE = 0.55$, $t = 3.03$, $p = .002$), suggesting a difference in the scalp topographies between social and biological implausibility effects. Separate analyses showed a significant interaction between Plausibility and Posteriority for biological utterances ($\beta = 1.31$, $SE = 0.40$, $t = 3.27$, $p = .001$), with larger implausibility effects over posterior sites (1.45 μV) than over anterior sites (0.16 μV). However, this interaction was not significant for social utterances ($\beta = -0.38$, $SE = 0.38$, $t = -0.99$, $p = .321$), suggesting comparable implausibility effects between posterior (−0.98 μV) and anterior sites (−0.60 μV).

Similarly, to determine whether implausibility effects differed between left and right sites, we conducted a laterality analysis. We fit the models with Plausibility, Type, and Laterality (left = −0.5, right = 0.5) as interacting fixed effects (see Table S3 in the Supplemental Material for model structures). Again, our focus was on the three-way interaction among Plausibility, Type, and Laterality. In the time window of 300–600 msec after the critical word onset, the interaction among Plausibility, Type, and Laterality did not reach statistical significance ($\beta = 0.33$, $SE = 0.48$, $t = 0.68$, $p = .495$). Separate analyses of social and biological utterances showed no significant interaction between Plausibility and Laterality for either social utterances ($\beta = -0.11$, $SE = 0.33$, $t = -0.32$, $p = .749$) or biological utterances ($\beta = 0.22$, $SE = 0.34$, $t = 0.64$, $p = .523$), suggesting comparable implausibility effects over left and right sites for both social utterances (−0.51 μV vs. −0.61 μV) and biological utterances (−0.01 μV vs. 0.21 μV). In the time window of 600–1000 msec after the critical word onset, the interaction among Plausibility, Type, and Laterality did not reach statistical significance ($\beta = 0.13$, $SE = 0.56$, $t = 0.23$, $p = .819$). Separate analyses of social and biological utterances showed no significant interaction between Plausibility and Laterality for either social utterances ($\beta = 0.13$, $SE = 0.38$, $t = 0.34$, $p = .736$) or biological utterances ($\beta = 0.26$, $SE = 0.40$, $t = 0.63$, $p = .528$), suggesting comparable implausibility effects over left and right sites for both social utterances (−0.85 μV vs. −0.73 μV) and biological utterances (0.68 μV vs. 0.93 μV).

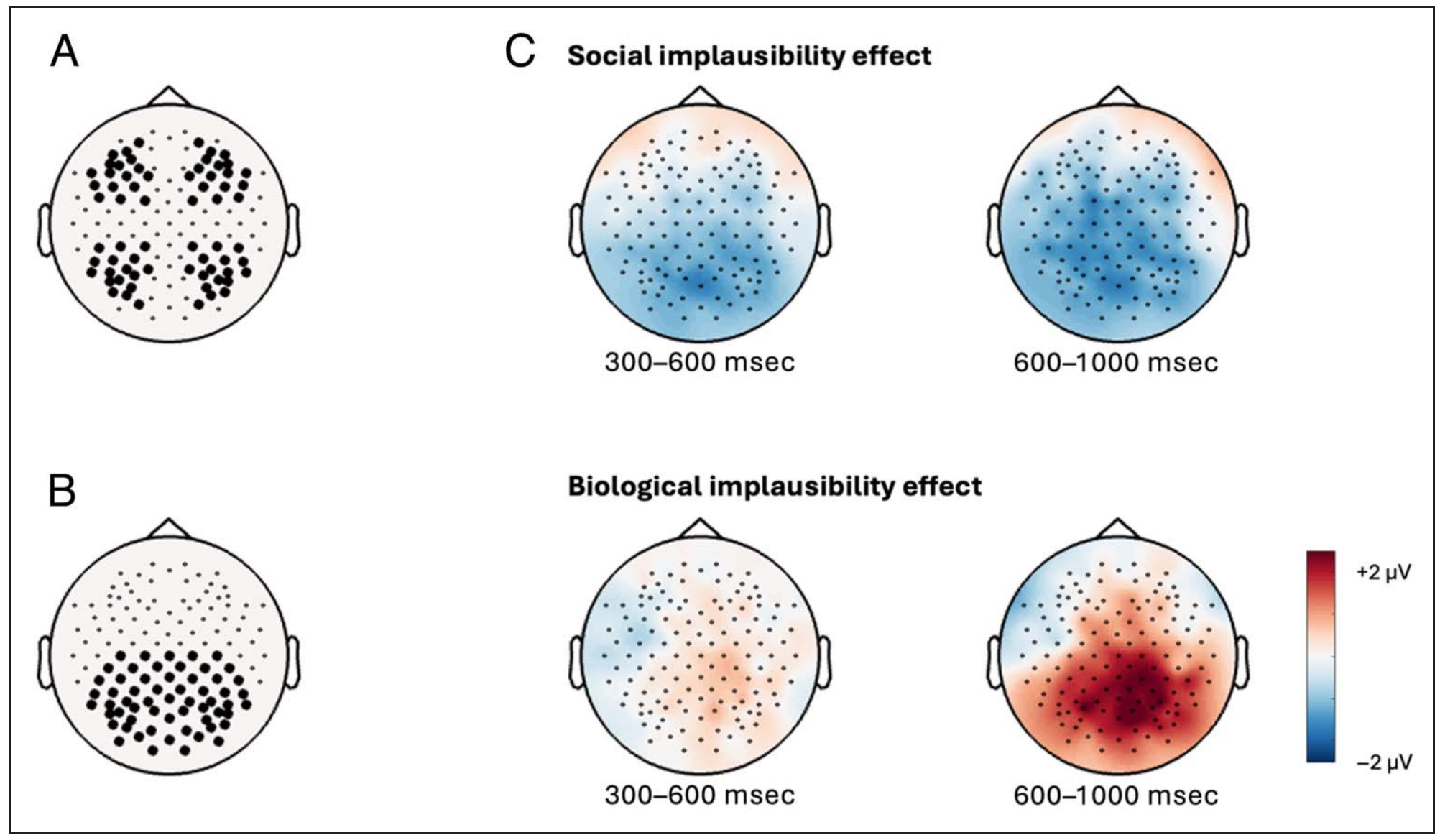

**Figure 3.** (A) Regions included in the topography analysis. (B) The ROI used in the main analyses. (C) Topographies of social and biological implausibility effects during the 300–600 and 600–1000 msec windows after the critical word onset.

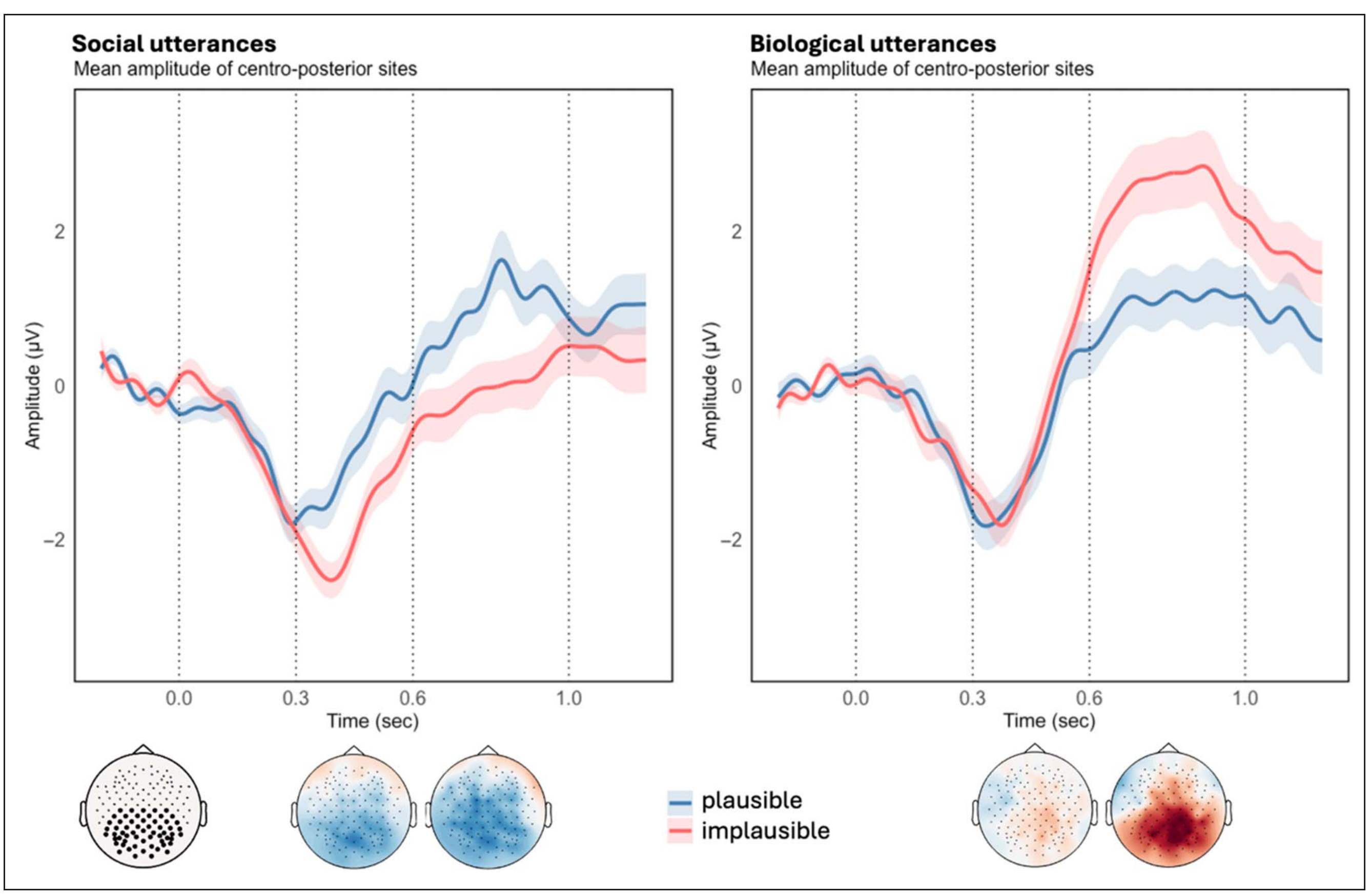

**Figure 4.** Grand-averaged ERPs for social and biological utterances. Analysis windows include 300–600 and 600–1000 msec post-critical word onset (0 msec). Shaded areas represent the *SEM* calculated across participants at each time point for each condition.

Overall, implausible utterances elicited distinct neurophysiological responses under social and biological conditions. Compared with plausible controls, socially implausible utterances elicited a larger negative deflection, while biologically implausible utterances elicited a larger positive deflection. As shown in Figure 3C, both social and biological implausibility effects were more pronounced over posterior sites than anterior sites, while no particular lateral asymmetry was found. These results aligned with the classic centro-posterior distribution of N400 and P600 effects in spoken language comprehension.

### Main Analyses

Based on the topographies of social and biological implausibility effects, we selected an ROI (Figure 3B) consisting of 59 centro-posterior sites (see Table S5 in the Supplemental Material for the full list) with mean amplitudes collapsed within the ROI. As complementary analyses, we additionally included channel-level analysis, in which the data of each channel were preserved (instead of being collapsed) and channel was included as a random effect in the LME models.

Our primary focus was to compare listeners' neural responses to social and biological implausibilities. We fit LME models with Plausibility and Type as interacting fixed effects for the mean amplitude over 300–600 and 600–1000 msec windows after the critical word onset (see Table S4 in the Supplemental Material for model structures). As shown in Figure 4, the results revealed a significant interaction between Plausibility and Type over 300–600 msec ($\beta = 1.16$, $SE = 0.50$, $t = 2.32$, $p = .021$) and over 600–1000 msec ($\beta = 2.44$, $SE = 0.59$, $t = 4.14$, $p < .001$). The further channel-level analysis confirmed these interactions (300–600 msec: $\beta = 1.17$, $SE = 0.52$, $t = 2.28$, $p = .026$; 600–1000 msec: $\beta = 2.46$, $SE = 0.59$, $t = 4.14$, $p < .001$).

For the 300–600 msec window, separate analyses showed a significant main effect of Plausibility for social utterances ($\beta = -0.89$, $SE = 0.32$, $t = -2.80$, $p = .005$) but no effect of Plausibility for biological utterances ($\beta = 0.27$, $SE = 0.34$, $t = 0.81$, $p = .419$), suggesting that socially implausible utterances elicited an N400 effect compared with their plausible control, which was not observed for biologically implausible utterances. The further channel-level analysis confirmed the significant N400 effect for social utterances ($\beta = -0.91$, $SE = 0.05$, $t = -18.99$, $p < .001$). This complementary analysis also revealed a significant, albeit small positive effect for biological utterances ($\beta = 0.29$, $SE =$ 0.05, $t = 5.68$, $p < .001$), suggesting that part of the P600 effect might have earlier onset.

For the 600–1000 msec window, separate analyses showed significant main effects of Plausibility for both social utterances ($\beta = -1.01$, $SE = 0.37$, $t = -2.75$, $p = .006$) and biological utterances ($\beta = 1.43$, $SE = 0.54$, $t = 2.66$, $p = .011$), suggesting that the N400 effect elicited by socially implausible utterances extended into the later time window. In contrast, biologically implausible utterances elicited a P600 effect compared with biologically plausible controls. The further channel-level analysis confirmed the extended N400 for social utterances ($\beta = -1.00$, $SE = 0.06$, $t = -17.74$, $p < .001$) and the P600 effect for biological utterances ($\beta = 1.44$, $SE = 0.66$, $t = 2.19$, $p = .031$).

We then examined whether social and biological implausibility effects were predicted by listeners' personality traits of openness by fitting LME models with Plausibility, Type, and Openness as interacting fixed effects (see Table S4 in the Supplemental Material for model structures). The results showed a significant three-way interaction among Plausibility, Type, and Openness during 300–600 msec ($\beta = -1.12$, $SE = 0.46$, $t = -2.41$, $p = .016$), but not during 600–1000 msec ($\beta = -0.88$, $SE = 0.53$, $t = -1.65$, $p = .099$). The further channel-level analysis confirmed the significant three-way interaction in the 300–600 msec window ($\beta = -1.07$, $SE = 0.07$, $t = -15.36$, $p < .001$) and also revealed a significant three-way interaction in the 600–1000 msec window ($\beta = -0.85$, $SE = 0.08$, $t = -10.42$, $p < .001$). Furthermore, the channel-level analysis revealed significant two-way interactions between Type and Openness in both time windows (300–600 msec: $\beta = -0.08$, $SE = 0.03$, $t = -2.26$, $p = .024$; 600–1000 msec: $\beta = -0.38$, $SE = 0.04$, $t = -9.38$, $p < .001$), which were not significant in the mean amplitude analysis.

Separate analyses of social and biological utterances showed that Plausibility significantly interacted with Openness for social utterances during 300–600 msec ($\beta = 0.71$, $SE = 0.32$, $t = 2.24$, $p = .026$), but not during 600–1000 msec ($\beta = 0.51$, $SE = 0.37$, $t = 1.38$, $p = .167$). This suggested that the social implausibility effect of N400 decreased as a function of a listener's score of openness. In contrast, this interaction did not reach statistical significance for biological utterances during either 300–600 msec ($\beta = -0.40$, $SE = 0.34$, $t = -1.16$, $p = .245$) or 600–1000 msec ($\beta = -0.36$, $SE = 0.48$, $t = -0.75$, $p = .459$), showing no evidence for the impact of a listeners' openness scores on the biological implausibility effect (Figure 5).

The further channel-level analysis largely confirmed this pattern and showed extra details. For social utterances, the interaction between plausibility and openness was confirmed in the earlier time window (300–600 msec, $\beta = 0.74$, $SE = 0.05$, $t = 15.23$, $p < .001$). However, it revealed an additional significant interaction in the later time window (600–1000 msec, $\beta = 0.60$, $SE = 0.06$, $t = 10.54$, $p < .001$), suggesting that openness also modulated the neural response to social implausibility in this later time window as the long-lasting integration effort continued. For biological utterances, the channel-level analysis revealed an interaction between plausibility and openness during the earlier time window (300–600 msec, $\beta = -0.33$, $SE = 0.05$, $t = -6.47$, $p < .001$), suggesting that participants with higher openness initially tried to integrate (a more negative implausibility effect as openness increased) even some biological violations in this early window.

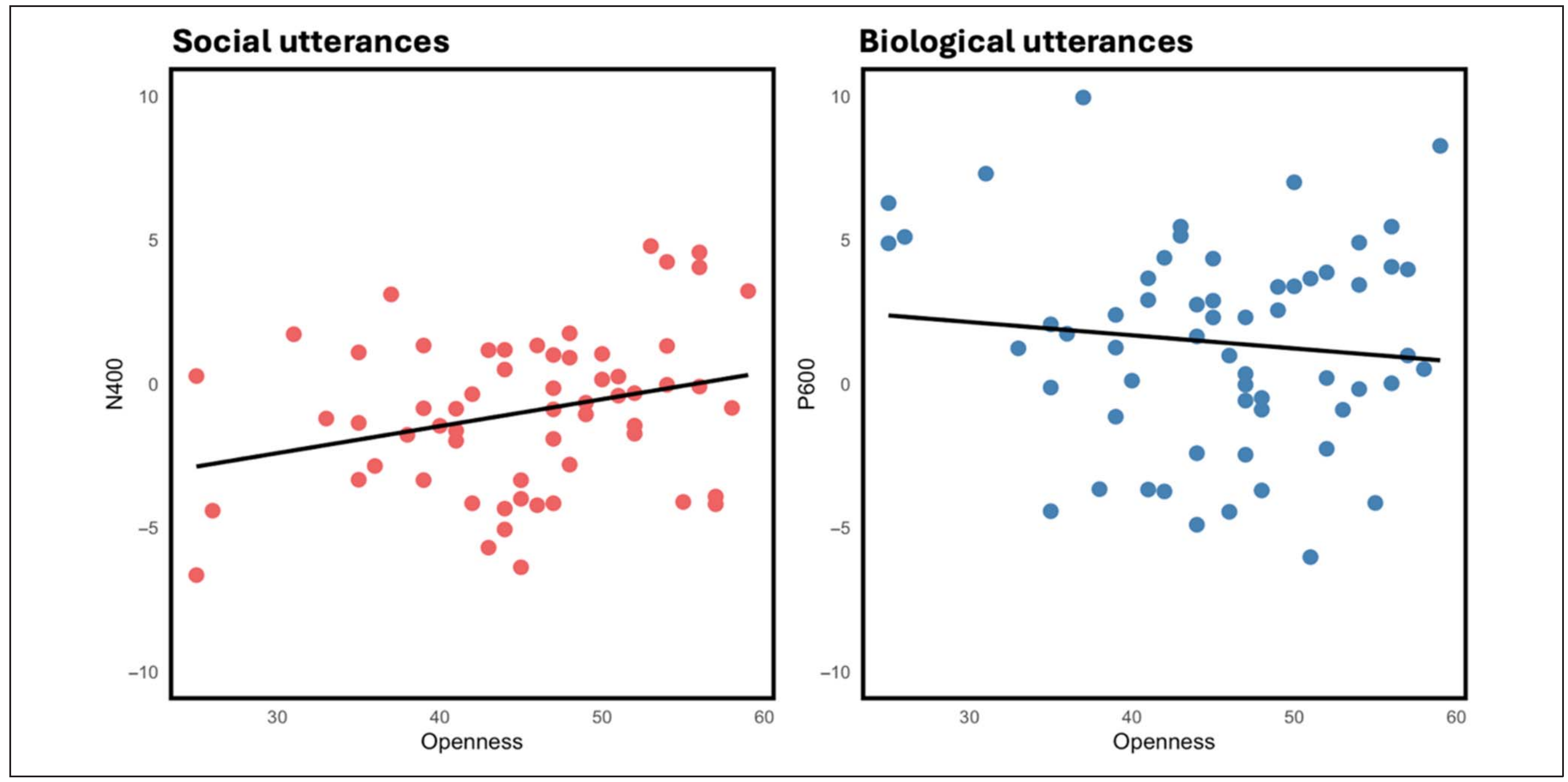

**Figure 5.** Social implausibility effect (N400) and biological implausibility effect (P600) predicted by openness scores.

Overall, social and biological implausibilities elicited distinct neural responses. Social implausibility elicited an N400 effect, while biological implausibility elicited a P600 effect. The social N400 effect was predicted by listeners' scores of openness, while the biological P600 effect was not.

### Further Analyses of Implausibility Effects

To examine whether the specific dimension of speaker identity (gender vs. age) differentially influenced neural responses to social and biological implausibility, we included Dimension (gender = −0.5, age = 0.5) as a fixed effect interacting with Plausibility and Type in our LME models. The results showed that the three-way interaction among Plausibility, Type, and Dimension was not significant during either 300–600 msec ($\beta = 0.37$, $SE = 1.02$, $t = 0.36$, $p = .720$) or 600–1000 msec ($\beta = 1.20$, $SE = 1.17$, $t = 1.02$, $p = .309$). In the further channel-level analysis, the three-way interaction among Plausibility, Type, and Dimension remained nonsignificant during both 300–600 msec ($\beta = 0.33$, $SE = 1.02$, $t = 0.32$, $p = .748$) and 600–1000 msec ($\beta = 1.19$, $SE = 1.19$, $t = 1.00$, $p = .319$). These results suggested no significant differences in neural responses between gender- and age-implausible utterances for the different types of implausibility, indicating that the observed N400/P600 effects were not primarily driven by the specific dimension (gender vs. age) used to create the mismatch, but likely by the nature (type) of the mismatch (social vs. biological).

As a further exploration, we leveraged the variability in the plausibility ratings of socially implausible utterances and examined whether the observed social N400 effect was modulated by the continuous degree of perceived implausibility. We used the plausibility ratings as a predictor for the amplitudes in the 300–600 msec window and 600–1000 msec window. The results showed that plausibility ratings for socially implausible utterances did not significantly predict the amplitudes during 300–600 msec ($\beta = -0.05$, $SE = 0.21$, $t = -0.24$, $p = .809$) or 600–1000 msec ($\beta = -0.16$, $SE = 0.27$, $t = -0.60$, $p = .551$). Complementary channel-level analyses also confirmed these null findings (300–600 msec: $\beta = -0.05$, $SE = 0.03$, $t = -1.70$, $p = .089$; 600–1000 msec: $\beta = -0.21$, $SE = 0.27$, $t = -0.77$, $p = .445$). In other words, within the category of socially implausible utterances, finer gradations of perceived implausibility seemed not to linearly modulate the N400 or P600 window amplitudes.

Finally, to determine the offset of the N400 and the P600 effects elicited by social and biological implausibilities, respectively, we conducted paired-sample $t$ tests on six consecutive 100-msec windows from 600 to 1200 msec after the critical word onset. The obtained $p$ values were adjusted using the false discovery rate method to reduce the chance of Type I error. The social N400 effect ended around 900 msec after the critical word onset (800–900 msec: $t(59) = -3.18$, adjusted $p = .013$; 900–1000 msec: $t(59) = -1.85$, adjusted $p = .103$), and the biological P600 effect ended around 1000 msec (900–1000 msec: $t(59) = 2.46$, adjusted $p = .025$; 1000–1100 msec: $t(59) = 1.81$, adjusted $p = .090$) after the critical word onset.

## DISCUSSION

Our study provides neurophysiological evidence for how the human brain employs distinct rational inference strategies to comprehend language within a speaker context. We demonstrated that the brain dynamically adapts its information-processing approach when faced with speaker–content mismatches, engaging different neurocognitive mechanisms depending on the nature and perceived reconcilability of the conflicting information. Using social and biological implausibilities as test cases, we showed that when an utterance violated social stereotypes but still permitted a literal, albeit effortful, interpretation (e.g., a man getting a manicure), an N400 effect emerged. This likely reflects a process of integrating unexpected but possible information into an existing mental model. Conversely, when an utterance violated more fundamental biological knowledge, rendering a literal interpretation highly implausible or impossible (e.g., a man getting pregnant), a P600 effect was observed. This suggests a shift to a different cognitive operation, likely involving error detection and a revision of the listener's understanding of the speaker or the linguistic content. Plausibility ratings confirmed that listeners perceived these two types of mismatches differently, with biological violations seen as far less plausible than social stereotype violation. The selective modulation of the N400 by the listener's openness trait for social implausibility, but not the P600 for biological implausibility, further supports this dissociation, as the N400 likely reflects a flexible integration process sensitive to individual differences in personality and cognitive style, while the P600 likely reflects a response to information that challenges common sense that constrains how people understand the principles of the biological world.

In our study, social implausibility elicited a long-lasting N400 effect until around 900 msec after the critical word onset. This offset aligns with previous studies showing a late offset of N400 elicited by speaker–content mismatches during auditory language comprehension (Pélissier & Ferragne, 2022; Martin et al., 2016). Besides the general longer duration of auditory N400 compared with visual N400 (Hagoort & Brown, 2000), this prolonged effect may reflect the extended time required to integrate multiple sources of information—the linguistic content and speaker identity.

It is also noteworthy that the specific dimension of speaker identity used to create these mismatches—gender (male vs. female) or age (adult vs. child)—did not significantly alter the pattern of neural responses. Our analyses including speaker dimension showed no significant interaction with plausibility and type for either the N400 or P600 time windows. This null finding suggests that the core distinction in processing social versus biological implausibilities and the engagement of correspondingly different inferential strategies reflected in the N400 and P600 effects are robust across speaker dimensions of age and gender.

### A Rational Account of Speaker-contextualized Language Comprehension

Our findings demonstrate how rational inference operates during real-time language processing. In Bayesian terms, listeners arrive at an interpretation (posterior) based on their world knowledge (prior) and their perception (evidence). The term "perception" here is twofold: On the one hand, it includes the listener's perception of the speaker's identity, which is formed rather swiftly upon hearing the speaker's voice (Lavan, Rinke, & Scharinger, 2024; McAleer et al., 2014; Scharinger et al., 2011); on the other hand, it includes the listener's perception of the linguistic content.

Upon hearing a socially implausible utterance, as social stereotypes appear to be "soft" constraints, listeners can make an effortful (as compared with the plausible controls) integration of their prior social stereotypes and the perceived evidence, resulting in an N400 effect. In contrast, hearing a biologically implausible utterance activates the biological knowledge that is rather "hard" constraints. The listener's perception substantially deviates from their prior biological knowledge, triggering an error correction process, either by revising their perception of the speaker or by revising the perceived linguistic content, and this correction process is reflected as a P600 effect.

It is valuable to consider our findings in light of frameworks linking ERPs to rational inference (e.g., Li & Ettinger, 2023; Ryskin et al., 2021). When an incoming word has a low prior probability given the context (including sentence and speaker information), comprehenders evaluate the likelihood that this perceived implausibility arose from an error (e.g., a speaker's slip of the tongue or a listener's mishearing). If the likelihood of an error is high, comprehenders may revise their perception toward a more plausible alternative, a process often associated with a P600 effect (and a reduced N400). If an error is not inferred despite the low plausibility of the literal input, an N400 effect may be more prominent and reflect the difficulty of integrating this surprising information.

Our plausibility rating data are consistent with this general perspective: Biologically implausible utterances were rated as far less plausible than socially implausible utterances. This stark difference in plausibility likely leads to a significantly higher estimated likelihood of an error for biological mismatches, thus guiding the cognitive system toward a P600-indexed error correction. Conversely, for social mismatches, the higher plausibility (relative to biological ones) likely results in a lower estimated likelihood of error, leading to an attempt at literal integration reflected by the N400.

It is worth noting that our account also presents a difference to models where ERPs simply track raw, continuous (im)plausibility. While the perceived (im)plausibility of an utterance is undoubtedly a strong contributor to this error likelihood estimation, implausibility and error likelihood are not necessarily identical. Other factors, such as the contextual availability of plausible alternatives, might also modulate the likelihood of error (see Li & Ettinger, 2023; Ryskin et al., 2021; Gibson et al., 2013). In our data, we found that within the category of socially implausible utterances, finer, continuous variations in their perceived implausibility did not significantly predict EEG amplitudes in either N400 or P600 window. While it is possible that the post hoc selection of only the socially implausible utterances might reduce the statistical power to detect effects, the null finding at least suggests against an account that neural responses are entirely determined by (the subtle gradations of) implausibility. Future research could aim for more precise, item-specific predictions by designing stimuli that directly manipulate the likelihood of error by manipulating not only the utterance plausibility but also other factors such as whether there are plausible alternatives, and what kinds of alternatives there are.

The rational inference account is further supported by our finding that a listener's openness trait generally modulates the social N400 effect but not the biological P600 effect. In rational inference, the integration process depends critically on the nature of prior knowledge—both its content and its distribution. For social stereotypes, prior knowledge varies substantially across individuals, with more open-minded individuals maintaining broader, less restrictive distributions of what they consider probable or acceptable for different social groups. Consequently, when encountering socially implausible utterances, listeners with higher openness levels show reduced N400 effects because the input deviates less from their more flexible prior expectations (as compared with listeners with lower openness levels). In contrast, biological knowledge represents more fundamental knowledge that remains relatively constant across individuals. When encountering biological implausibility, even the most open-minded listeners recognize the substantial deviation from the common sense of human biological constraints, triggering error correction processes reflected in the P600. This dissociation in how openness selectively modulates social but not biological implausibility processing aligns with what we would expect if these ERP components indeed reflect different aspects of rational inference—integration difficulty (N400) when prior knowledge allows for an interpretation, and error correction (P600) when the input substantially violates prior knowledge.

The complementary channel-level results provided more details regarding this rational inference process.

For example, there was a small but significant positive-going effect for biologically implausible utterances even in the earlier 300–600 msec window, suggesting part of the P600 effect might have earlier onset. Interestingly, the amplitude in this earlier window for biological implausibility was also modulated by listeners' openness: Individuals with lower openness tended to show this early positivity (indicative of quicker error detection/correction), whereas those with higher openness sometimes showed a more negative-going deflection (potentially reflecting an initial attempt at integrating even some biological mismatches). This suggests that while biological constraints are generally robust, highly open-minded individuals might initially try to integrate them before error correction mechanisms fully engage.

Additionally, these channel-level analyses revealed that openness also modulated the neural response to socially implausible utterances in the later 600–1000 msec window. This finding supports the interpretation that the social N400 effect, reflecting effortful integration, was indeed long lasting and that this extended integration process continued to be influenced by openness.

It should be noted that biological implausibility, as a trigger for P600 effects, only represents one typical case where error correction is required. It does not exclude other cases that may require similar processing and thus trigger P600 effects. For instance, in the study by Lattner and Friederici (2003), the absence of fillers or the limited number of speakers might have led listeners to stop integrating social stereotypes and switch to a strategic error detection process after several trials, resulting in a P600 effect (see also Foucart et al., 2015; Van Berkum et al., 2008, for discussion).

## The Next Questions

Our findings raise questions for future research. While we propose an error correction mechanism for the P600 observed for biological implausibility, the exact nature of these "errors" and their correction process remains an open question. We suggest two potential processes: Listeners may either revise the perceived linguistic content to align with the speaker's identity or adjust the perceived speaker identity. Revision of perceived linguistic content has been demonstrated in Cai et al. (2022), where they showed that listeners estimated the likelihood of an implausible sentence as being corrupted from an otherwise plausible one and sometimes revised the implausible sentence into a plausible sentence (e.g., revising "The mother gave the candle the daughter" into "The mother gave the candle *to* the daughter") to arrive at a plausible interpretation (see also Gibson et al., 2013). While our sentences are semantically plausible in isolation, listeners might still infer that the speaker misspoke a critical word or intended a different but related concept. Alternatively, another potential process is an adjustment of the perceived speaker identity. Listeners might reevaluate their initial assessment of the speaker identity for that specific utterance (e.g., a voice initially perceived as a male voice might be actually from a female speaker with a lower voice). At present, both types of revision—of the linguistic content or of the speaker identity—appear to be plausible mechanisms for resolving the incongruity in biologically implausible scenarios. Future research can design specific stimuli to distinguish under what conditions one strategy might be favored over the other or how they might interact.

Future research can also explore broader personal factors in terms of their influence on rational inference as a part of social cognition. As language fundamentally serves social communication, investigating how various personal factors—from personality traits to political affiliations, education level, and culture—affect inference processes could provide valuable insights into the interface between language and social cognition. Furthermore, our participant sample was not characterized by gender identity beyond male/female categories, and our gender-related stimuli were constructed based on a common binary gender assumption. Future research is encouraged to include participants with diverse gender identities, sexual orientations, or unique experiences and perspectives to explore how listeners' characteristics might interact with speaker characteristics. For example, the interpretation of what constitutes a "social" versus a "biological" mismatch might vary within these communities. In our current study, we also manipulated speaker identity using an age dimension (child vs. adult) in addition to gender, and our speaker dimension analysis did not reveal significant differences in the core N400/P600 dissociation for social versus biological mismatches across these two dimensions for our current sample. While this finding might suggest generality in the rational inference processes triggered by these mismatch types, future research can investigate how these processes operate within more diverse populations.

## Conclusion

Our study provides neural evidence for adaptive rational inference processes in real-time language comprehension and social cognition. The brain implements distinct approaches—effortful integration (N400) versus error correction (P600)—depending on whether incoming linguistic information, contextualized by the speaker's identity, challenges listeners' expectation formed by malleable social stereotypes or more immutable biological knowledge. This dissociation highlights the brain's adaptive information-processing capabilities, where distinct neural responses are recruited to manage different types of conflict between incoming information and internal representations. These findings demonstrate how the brain flexibly and rationally constructs meaning in dynamic, socially embedded communication, reconciling conflicting information from multiple sources to achieve coherent understanding.

## Acknowledgments

We thank Xiaohui Rao for her assistance during data collection and discussion. We thank Pengcheng Wang for his insights and helpful discussion.

Corresponding author: Zhenguang G. Cai, Department of Linguistics and Modern Languages, The Chinese University of Hong Kong, Hong Kong, SAR, e-mail: zhenguangcai@cuhk.edu.hk.

## Data Availability Statement

The experimental stimuli, data, and code of this study are available at https://osf.io/dhtxn/. Supplemental Material can be accessed on this article's homepage: https://doi.org/10.1162/JOCN.a.102.

## Author Contributions

Hanlin Wu: Conceptualization; Data curation; Formal analysis; Investigation; Methodology; Resources; Software; Visualization; Writing—Original draft; Writing—Review & editing. Zhenguang G. Cai: Conceptualization; Funding acquisition; Supervision; Writing—Original draft; Writing—Review & editing.

## Funding Information

This work was supported by the General Research Fund (https://dx.doi.org/10.13039/501100002920; grant no. 14600220), University Grants Committee, Hong Kong.

## Diversity in Citation Practices

Retrospective analysis of the citations in every article published in this journal from 2010 to 2021 reveals a persistent pattern of gender imbalance: Although the proportions of authorship teams (categorized by estimated gender identification of first author/last author) publishing in the *Journal of Cognitive Neuroscience* (*JoCN*) during this period were M(an)/M = .407, W(oman)/M = .32, M/W = .115, and W/W = .159, the comparable proportions for the articles that these authorship teams cited were M/M = .549, W/M = .257, M/W = .109, and W/W = .085 (Postle and Fulvio, *JoCN*, 34:1, pp. 1–3). Consequently, *JoCN* encourages all authors to consider gender balance explicitly when selecting which articles to cite and gives them the opportunity to report their article's gender citation balance. The authors of this article report its proportions of citations by gender category to be: M/M = .424; W/M = .288; M/W = .102; W/W = .186.

## REFERENCES

Bornkessel-Schlesewsky, I., Krauspenhaar, S., & Schlesewsky, M. (2013). Yes, you can? A speaker's potency to act upon his words orchestrates early neural responses to message-level meaning. *PLoS One*, *8*, e69173. https://doi.org/10.1371/journal.pone.0069173, PubMed: 23894425

Brothers, T., Dave, S., Hoversten, L. J., Traxler, M. J., & Swaab, T. Y. (2019). Flexible predictions during listening comprehension: Speaker reliability affects anticipatory processes. *Neuropsychologia*, *135*, 107225. https://doi.org/10.1016/j.neuropsychologia.2019.107225, PubMed: 31605686

Brown-Schmidt, S., Yoon, S. O., & Ryskin, R. A. (2015). People as contexts in conversation. In *Psychology of learning and motivation—Advances in research and theory* (Vol. 62, pp. 59–99). Elsevier. https://doi.org/10.1016/bs.plm.2014.09.003

Caffarra, S., Wolpert, M., Scarinci, D., & Mancini, S. (2020). Who are you talking to? The role of addressee identity in utterance comprehension. *Psychophysiology*, *57*, e13527. https://doi.org/10.1111/psyp.13527, PubMed: 31953848

Cai, Q., & Brysbaert, M. (2010). SUBTLEX-CH: Chinese word and character frequencies based on film subtitles. *PLoS One*, *5*, e10729. https://doi.org/10.1371/journal.pone.0010729, PubMed: 20532192

Cai, Z. G. (2022). Interlocutor modelling in comprehending speech from interleaved interlocutors of different dialectic backgrounds. *Psychonomic Bulletin & Review*, *29*, 1026–1034. https://doi.org/10.3758/s13423-022-02055-7, PubMed: 35106731

Cai, Z. G., Gilbert, R. A., Davis, M. H., Gaskell, M. G., Farrar, L., Adler, S., et al. (2017). Accent modulates access to word meaning: Evidence for a speaker-model account of spoken word recognition. *Cognitive Psychology*, *98*, 73–101. https://doi.org/10.1016/j.cogpsych.2017.08.003, PubMed: 28881224

Cai, Z. G., Zhao, N., & Pickering, M. J. (2022). How do people interpret implausible sentences? *Cognition*, *225*, 105101. https://doi.org/10.1016/j.cognition.2022.105101, PubMed: 35339795

Chen, P. G., & Palmer, C. L. (2018). The prejudiced personality? Using the big five to predict susceptibility to stereotyping behavior. *American Politics Research*, *46*, 276–307. https://doi.org/10.1177/1532673X17719720

Crawford, J. T., & Brandt, M. J. (2019). Who is prejudiced, and toward whom? The big five traits and generalized prejudice. *Personality and Social Psychology Bulletin*, *45*, 1455–1467. https://doi.org/10.1177/0146167219832335, PubMed: 30895844

Creel, S. C., & Bregman, M. R. (2011). How talker identity relates to language processing. *Language and Linguistics Compass*, *5*, 190–204. https://doi.org/10.1111/J.1749-818X.2011.00276.X

Digman, J. M. (1990). Personality structure: Emergence of the five-factor model. *Annual Review of Psychology*, *41*, 417–440. https://doi.org/10.1146/annurev.ps.41.020190.002221

Flynn, F. J. (2005). Having an open mind: The impact of openness to experience on interracial attitudes and impression formation. *Journal of Personality and Social Psychology*, *88*, 816–826. https://doi.org/10.1037/0022-3514.88.5.816, PubMed: 15898877

Foucart, A., Garcia, X., Ayguasanosa, M., Thierry, G., Martin, C., & Costa, A. (2015). Does the speaker matter? Online processing of semantic and pragmatic information in L2 speech comprehension. *Neuropsychologia*, *75*, 291–303. https://doi.org/10.1016/j.neuropsychologia.2015.06.027, PubMed: 26115602

Foucart, A., & Hartsuiker, R. J. (2021). Are foreign-accented speakers that 'incredible'? The impact of the speaker's indexical properties on sentence processing. *Neuropsychologia*, *158*, 107902. https://doi.org/10.1016/j.neuropsychologia.2021.107902, PubMed: 34052231

Foucart, A., Santamaría-García, H., & Hartsuiker, R. J. (2019). Short exposure to a foreign accent impacts subsequent

cognitive processes. *Neuropsychologia*, *129*, 1–9. https://doi.org/10.1016/j.neuropsychologia.2019.02.021, PubMed: 30858062

Frömer, R., Maier, M., & Abdel Rahman, R. (2018). Group-level EEG-processing pipeline for flexible single trial-based analyses including linear mixed models. *Frontiers in Neuroscience*, *12*, 48. https://doi.org/10.3389/fnins.2018.00048, PubMed: 29472836

Gibson, E., Bergen, L., & Piantadosi, S. T. (2013). Rational integration of noisy evidence and prior semantic expectations in sentence interpretation. *Proceedings of the National Academy of Sciences, U.S.A.*, *110*, 8051–8056. https://doi.org/10.1073/pnas.1216438110, PubMed: 23637344

Grant, A., Grey, S., & van Hell, J. G. (2020). Male fashionistas and female football fans: Gender stereotypes affect neurophysiological correlates of semantic processing during speech comprehension. *Journal of Neurolinguistics*, *53*, 100876. https://doi.org/10.1016/j.jneuroling.2019.100876

Hagoort, P., Baggio, G., & Willems, R. M. (2009). Semantic unification. In M. S. Gazzaniga (Ed.), *The cognitive neurosciences* (4th ed., pp. 819–836). Cambridge, MA: MIT Press. https://doi.org/10.7551/mitpress/8029.003.0072

Hagoort, P., Brown, C., & Groothusen, J. (1993). The syntactic positive shift (SPS) as an ERP measure of syntactic processing. *Language and Cognitive Processes*, *8*, 439–483. https://doi.org/10.1080/01690969308407585

Hagoort, P., & Brown, C. M. (2000). ERP effects of listening to speech: Semantic ERP effects. *Neuropsychologia*, *38*, 1518–1530. https://doi.org/10.1016/S0028-3932(00)00052-X, PubMed: 10906377

Hagoort, P., Hald, L., Bastiaansen, M., & Petersson, K. M. (2004). Integration of word meaning and world knowledge in language comprehension. *Science*, *304*, 438–441. https://doi.org/10.1126/science.1095455, PubMed: 15031438

Hanulíková, A., & Carreiras, M. (2015). Electrophysiology of subject–verb agreement mediated by speakers' gender. *Frontiers in Psychology*, *6*, 1396. https://doi.org/10.3389/fpsyg.2015.01396, PubMed: 26441771

Hanulíková, A., van Alphen, P. M., van Goch, M. M., & Weber, A. (2012). When one person's mistake is another's standard usage: The effect of foreign accent on syntactic processing. *Journal of Cognitive Neuroscience*, *24*, 878–887. https://doi.org/10.1162/jocn_a_00103, PubMed: 21812565

Hay, J., Warren, P., & Drager, K. (2006). Factors influencing speech perception in the context of a merger-in-progress. *Journal of Phonetics*, *34*, 458–484. https://doi.org/10.1016/j.wocn.2005.10.001

Heise, M. J., Mon, S. K., & Bowman, L. C. (2022). Utility of linear mixed effects models for event-related potential research with infants and children. *Developmental Cognitive Neuroscience*, *54*, 101070. https://doi.org/10.1016/j.dcn.2022.101070, PubMed: 35395594

Hernández-Gutierrez, D., Muñoz, F., Sánchez-García, J., Sommer, W., Abdel Rahman, R., Casado, P., et al. (2021). Situating language in a minimal social context: How seeing a picture of the speaker's face affects language comprehension. *Social Cognitive and Affective Neuroscience*, *16*, 502–511. https://doi.org/10.1093/scan/nsab009, PubMed: 33470410

Johnson, K., Strand, E. A., & D'Imperio, M. (1999). Auditory-visual integration of talker gender in vowel perception. *Journal of Phonetics*, *27*, 359–384. https://doi.org/10.1006/jpho.1999.0100

Kemmerer, D., Weber-Fox, C., Price, K., Zdanczyk, C., & Way, H. (2007). Big brown dog or brown big dog? An electrophysiological study of semantic constraints on prenominal adjective order. *Brain and Language*, *100*, 238–256. https://doi.org/10.1016/j.bandl.2005.12.002, PubMed: 16412501

Kim, A., & Osterhout, L. (2005). The independence of combinatory semantic processing: Evidence from event-related potentials. *Journal of Memory and Language*, *52*, 205–225. https://doi.org/10.1016/j.jml.2004.10.002

Kim, J. (2016). Perceptual associations between words and speaker age. *Laboratory Phonology*, *7*, 18. https://doi.org/10.5334/labphon.33

Kutas, M., & Hillyard, S. A. (1980). Reading senseless sentences: Brain potentials reflect semantic incongruity. *Science*, *207*, 203–205. https://doi.org/10.1126/science.7350657, PubMed: 7350657

Kutas, M., & Hillyard, S. A. (1984). Brain potentials during reading reflect word expectancy and semantic association. *Nature*, *307*, 161–163. https://doi.org/10.1038/307161a0, PubMed: 6690995

Lattner, S., & Friederici, A. D. (2003). Talker's voice and gender stereotype in human auditory sentence processing—Evidence from event-related brain potentials. *Neuroscience Letters*, *339*, 191–194. https://doi.org/10.1016/S0304-3940(03)00027-2, PubMed: 12633885

Lavan, N., Rinke, P., & Scharinger, M. (2024). The time course of person perception from voices in the brain. *Proceedings of the National Academy of Sciences, U.S.A.*, *121*, e2318361121. https://doi.org/10.1073/pnas.2318361121, PubMed: 38889147

Levy, R. (2008). A noisy-channel model of rational human sentence comprehension under uncertain input. In *EMNLP '08: Proceedings of the Conference on Empirical Methods in Natural Language Processing* (pp. 234–243). https://doi.org/10.3115/1613715.1613749

Li, J., & Ettinger, A. (2023). Heuristic interpretation as rational inference: A computational model of the N400 and P600 in language processing. *Cognition*, *233*, 105359. https://doi.org/10.1016/j.cognition.2022.105359, PubMed: 36549129

Luck, S. J. (2014). *An introduction to the event-related potential technique* (2nd ed.). MIT Press. https://books.google.com.hk/books?id=y4-uAwAAQBAJ

Luck, S. J. (2022). *Applied event-related potential data analysis*. https://socialsci.libretexts.org/Bookshelves/Psychology/Book%3A_Applied_Event-Related_Potential_Data_Analysis_(Luck)

Martin, C. D., Garcia, X., Potter, D., Melinger, A., & Costa, A. (2016). *Holiday* or *vacation*? The processing of variation in vocabulary across dialects. *Language, Cognition and Neuroscience*, *31*, 375–390. https://doi.org/10.1080/23273798.2015.1100750

Matuschek, H., Kliegl, R., Vasishth, S., Baayen, H., & Bates, D. (2017). Balancing Type I error and power in linear mixed models. *Journal of Memory and Language*, *94*, 305–315. https://doi.org/10.1016/j.jml.2017.01.001

McAleer, P., Todorov, A., & Belin, P. (2014). How do you say "hello"? Personality impressions from brief novel voices. *PLoS One*, *9*, e90779. https://doi.org/10.1371/journal.pone.0090779, PubMed: 24622283

Münte, T. F., Heinze, H. J., Matzke, M., Wieringa, B. M., & Johannes, S. (1998). Brain potentials and syntactic violations revisited: No evidence for specificity of the syntactic positive shift. *Neuropsychologia*, *36*, 217–226. https://doi.org/10.1016/S0028-3932(97)00119-X, PubMed: 9622187

Nieuwland, M. S., & Van Berkum, J. J. A. (2006). When peanuts fall in love: N400 evidence for the power of discourse. *Journal of Cognitive Neuroscience*, *18*, 1098–1111. https://doi.org/10.1162/jocn.2006.18.7.1098, PubMed: 16839284

Oostenveld, R., Fries, P., Maris, E., & Schoffelen, J.-M. (2011). FieldTrip: Open source software for advanced analysis of MEG, EEG, and invasive electrophysiological data. *Computational Intelligence and Neuroscience*, *2011*,

156869. https://doi.org/10.1155/2011/156869, PubMed: 21253357

Osterhout, L., & Holcomb, P. J. (1992). Event-related brain potentials elicited by syntactic anomaly. *Journal of Memory and Language*, *31*, 785–806. https://doi.org/10.1016/0749-596X(92)90039-Z

Pélissier, M., & Ferragne, E. (2022). The N400 reveals implicit accent-induced prejudice. *Speech Communication*, *137*, 114–126. https://doi.org/10.1016/j.specom.2021.10.004

Regel, S., Coulson, S., & Gunter, T. C. (2010). The communicative style of a speaker can affect language comprehension? ERP evidence from the comprehension of irony. *Brain Research*, *1311*, 121–135. https://doi.org/10.1016/j.brainres.2009.10.077, PubMed: 19900421

Ryskin, R., Ng, S., Mimnaugh, K., Brown-Schmidt, S., & Federmeier, K. D. (2020). Talker-specific predictions during language processing. *Language, Cognition and Neuroscience*, *35*, 797–812. https://doi.org/10.1080/23273798.2019.1630654, PubMed: 33693050

Ryskin, R., Stearns, L., Bergen, L., Eddy, M., Fedorenko, E., & Gibson, E. (2021). An ERP index of real-time error correction within a noisy-channel framework of human communication. *Neuropsychologia*, *158*, 107855. https://doi.org/10.1016/j.neuropsychologia.2021.107855, PubMed: 33865848

Scharinger, M., Monahan, P. J., & Idsardi, W. J. (2011). You had me at "Hello": Rapid extraction of dialect information from spoken words. *Neuroimage*, *56*, 2329–2338. https://doi.org/10.1016/j.neuroimage.2011.04.007, PubMed: 21511041

Scott, S. K. (2019). From speech and talkers to the social world: The neural processing of human spoken language. *Science*, *366*, 58–62. https://doi.org/10.1126/science.aax0288, PubMed: 31604302

Sibley, C. G., & Duckitt, J. (2008). Personality and prejudice: A meta-analysis and theoretical review. *Personality and Social Psychology Review*, *12*, 248–279. https://doi.org/10.1177/1088868308319226, PubMed: 18641385

Soto, C. J., & John, O. P. (2017). The next Big Five Inventory (BFI-2): Developing and assessing a hierarchical model with 15 facets to enhance bandwidth, fidelity, and predictive power. *Journal of Personality and Social Psychology*, *113*, 117–143. https://doi.org/10.1037/pspp0000096, PubMed: 27055049

Tanner, D., Morgan-Short, K., & Luck, S. J. (2015). How inappropriate high-pass filters can produce artifactual effects and incorrect conclusions in ERP studies of language and cognition. *Psychophysiology*, *52*, 997–1009. https://doi.org/10.1111/psyp.12437, PubMed: 25903295

Van Berkum, J. J. A., van den Brink, D., Tesink, C. M. J. Y., Kos, M., & Hagoort, P. (2008). The neural integration of speaker and message. *Journal of Cognitive Neuroscience*, *20*, 580–591. https://doi.org/10.1162/jocn.2008.20054, PubMed: 18052777

van de Meerendonk, N., Kolk, H. H. J., Chwilla, D. J., & Vissers, C. T. W. M. (2009). Monitoring in language perception. *Language and Linguistics Compass*, *3*, 1211–1224. https://doi.org/10.1111/j.1749-818X.2009.00163.x

van den Brink, D., Van Berkum, J. J. A., Bastiaansen, M. C. M., Tesink, C. M. J. Y., Kos, M., Buitelaar, J. K., et al. (2012). Empathy matters: ERP evidence for inter-individual differences in social language processing. *Social Cognitive and Affective Neuroscience*, *7*, 173–183. https://doi.org/10.1093/scan/nsq094, PubMed: 21148175

Walker, A., & Hay, J. (2011). Congruence between 'word age' and 'voice age' facilitates lexical access. *Laboratory Phonology*, *2*, 219–237. https://doi.org/10.1515/labphon.2011.007

Willems, R. M., & Hagoort, P. (2007). Neural evidence for the interplay between language, gesture, and action: A review. *Brain and Language*, *101*, 278–289. https://doi.org/10.1016/j.bandl.2007.03.004, PubMed: 17416411

Willems, R. M., Özyürek, A., & Hagoort, P. (2008). Seeing and hearing meaning: ERP and fMRI evidence of word versus picture integration into a sentence context. *Journal of Cognitive Neuroscience*, *20*, 1235–1249. https://doi.org/10.1162/jocn.2008.20085, PubMed: 18284352

Winkler, I., Haufe, S., & Tangermann, M. (2011). Automatic classification of artifactual ICA-components for artifact removal in EEG signals. *Behavioral and Brain Functions*, *7*, 30. https://doi.org/10.1186/1744-9081-7-30, PubMed: 21810266

Wu, H., & Cai, Z. G. (2024). Speaker effects in spoken language comprehension. *arXiv*. https://doi.org/10.48550/arXiv.2412.07238

Wu, H., Duan, X., & Cai, Z. G. (2024). Speaker demographics modulate listeners' neural correlates of spoken word processing. *Journal of Cognitive Neuroscience*, *36*, 2208–2226. https://doi.org/10.1162/jocn_a_02225, PubMed: 39023368

Zhang, B., Li, Y. M., Li, J., Luo, J., Ye, Y., Yin, L., et al. (2022). The Big Five Inventory-2 in China: A comprehensive psychometric evaluation in four diverse samples. *Assessment*, *29*, 1262–1284. https://doi.org/10.1177/10731911211008245, PubMed: 33884926

Zhou, P., Garnsey, S., & Christianson, K. (2019). Is imagining a voice like listening to it? Evidence from ERPs. *Cognition*, *182*, 227–241. https://doi.org/10.1016/j.cognition.2018.10.014, PubMed: 30366220